\documentclass[11pt]{article}

\usepackage{amsmath}
\usepackage{amssymb}
\usepackage{bbm}
\usepackage{bbold}
\usepackage{braket}
\usepackage{caption}
\usepackage{cite}
\usepackage{color}
    \definecolor{darkgreen}{rgb}{0,0.5,0}
    \definecolor{darkred}{rgb}{0.5,0,0}
    \definecolor{darkblue}{rgb}{0,0,0.6}
    \definecolor{purple}{rgb}{0.4,.2,0.7}
\usepackage[mathcal]{eucal}
\usepackage{float}
\usepackage{graphicx}
\usepackage[hyperfootnotes = true, colorlinks = true, linkcolor = darkblue, citecolor = purple]{hyperref}
\usepackage{mathtools}
\usepackage{pdfsync}
\usepackage{slashed}
\usepackage[normalem]{ulem}
\usepackage{upgreek}
\usepackage{url}
\allowdisplaybreaks

\usepackage[margin = 2.2cm]{geometry}
\usepackage[ragged]{footmisc}
\def\be{\begin{equation}}
\def\ee{\end{equation}}

\renewcommand{\tilde}{\widetilde}
\DeclareMathOperator{\Tr}{Tr}
\DeclareMathOperator{\arcsinh}{arcsinh}

\numberwithin{equation}{section}

\newcommand{\ben}{\begin{eqnarray}\displaystyle}
\newcommand{\een}{\end{eqnarray}}
\newcommand{\non}{\nonumber}
\newcommand{\refb}{\eqref}

\begin{document}

\thispagestyle{empty}
\begin{center}
    ~\vspace{5mm}
    
    {\Large \bf 
    
    Normalization of ZZ instanton amplitudes in minimal string theory
    
    }
    
    \vspace{0.4in}
    
    {\bf 
    Dan Stefan Eniceicu,$^1$ 
    Raghu Mahajan,$^1$
    Chitraang Murdia,$^{2,3}$ and 
    Ashoke Sen.$^{4}$
    }

    \vspace{0.4in}

    $^1$ Department of Physics, Stanford University, Stanford, CA 94305, USA \vskip1ex
    $^2$ Berkeley Center for Theoretical Physics, Department of Physics, University of California, Berkeley, CA 94720, USA \vskip1ex
    $^3$ Theoretical Physics Group, Lawrence Berkeley National Laboratory, Berkeley, CA 94720, USA \vskip1ex
    $^4$ International Centre for Theoretical Sciences, 
Bengaluru - 560089, India
    \vspace{0.1in}
    
    {\tt eniceicu@stanford.edu, raghumahajan@stanford.edu, murdia@berkeley.edu,
    ashoke.sen@icts.res.in}
\end{center}

\vspace{0.4in}

\begin{abstract}

We use insights from string field theory to analyze and cure the divergences in the cylinder diagram in minimal string theory with both boundaries lying on a ZZ brane.
We focus on theories with worldsheet matter consisting of the $(2,p)$ minimal model plus Liouville theory,  with total central charge 26,  together with the usual $bc$-ghosts.
The string field theory procedure gives a finite,  purely imaginary normalization constant for non-perturbative effects in minimal string theory,  or doubly non-perturbative effects in JT gravity.  
We find precise agreement with the prediction from the dual double-scaled one-matrix integral.
We also make a few remarks about the extension of this result to the more general $(p',p)$ minimal string.

\end{abstract}

\pagebreak

\tableofcontents

\section{Introduction and summary}
The duality between minimal string theory and double-scaled matrix integrals \cite{DiFrancesco:1993cyw} is the earliest known example of a duality between a gravitational and a non-gravitational system.
The term minimal string theory refers to two-dimensional gravity coupled to $c<1$ minimal models.
These are non-critical string theories where the Liouville mode does not decouple.
More precisely,  the worldsheet theory consists of the $(p',p)$ minimal model plus Liouville theory,  with total central charge 26,  together with the usual $bc$-ghosts.
Here $p'$ and $p$ are relatively-prime positive integers,  and we work with the convention that $p>p'\geq 2$.
The models with $(2,p)$ matter are dual to matrix integrals over just one matrix.
The $p \to \infty$ limit of the $(2,p)$ family is JT gravity \cite{sss},  a subject which has been of much recent interest.

While the minimal string theories are toy models,  one of the lessons from them that generalizes to even critical superstring theories is the existence of stronger-than-expected non-perturbative effects \cite{Shenker:1990uf,  PolchinskiCombinatorics}.
Let $g_s$ be the closed string coupling.
Then the non-perturbative effects are of order $\exp \left(- C g_s^{-1} \right)$, 
rather than the $\exp \left(- C g_s^{-2} \right)$ expected from field theory.
In the language of JT gravity, $g_s \propto e^{-S_0}$ where $S_0$ is the coefficient
of the Euler characteristic term in the action.
Given this identification,  these effects are ``doubly-nonperturbative" in the parameter $S_0$ \cite{sss}.

These non-perturbative effects are known to arise from ZZ branes on the string theory side \cite{zz}.
In the matrix integral,  these effects correspond to one-eigenvalue instantons.
A one-eigenvalue instanton refers to a subleading saddle point configuration in the matrix integral which differs from the leading saddle point by pulling one eigenvalue out of the droplet of eigenvalues and placing it at an extremum of the one-eigenvalue effective action \cite{Shenker:1990uf, David:1990sk}.

Let us consider the computation of the matrix integral $\mathfrak{Z}$ itself.
Let $T$ denote the action of the one-eigenvalue instanton or the ``tension" of the ZZ brane,  which is a positive quantity of order $g_s^{-1}$.
The quantity $\mathfrak{Z}$ admits an expansion of the form
\begin{align}
\mathfrak{Z} &= 
\mathfrak{Z}^{(0)} + \mathfrak{Z}^{(1)} + \ldots = 
\mathfrak{Z}^{(0)} \left( 1 + \mathcal{N} \, e^{-T} + \ldots \right) \, .
\label{zexpansion}
\end{align}
Here $\mathfrak{Z}^{(0)}$ is the perturbative contribution to the matrix integral and $\mathfrak{Z}^{(1)}$ is the contribution to the matrix integral when one eigenvalue is in the classically forbidden region.
One can also write (\ref{zexpansion}) in terms of the free energy as $\log \mathfrak{Z} = \log \mathfrak{Z}^{(0)} +  \mathcal{N} \, e^{-T}  + \ldots $.

The object of interest to us in this paper is the normalization constant $\mathcal{N}$.
Roughly speaking,  in string theory,  $\mathcal{N}$ is the exponential of the worldsheet annulus with ZZ boundary conditions on both ends.
This annulus amplitude has been computed using the worldsheet theory \cite{Martinec:2003ka,  seibergannulus} and is divergent.
However,  starting  with \cite{Silvestrov:1990ha,  David:1992za},  many papers have computed
a finite value for $\mathcal{N}$ using matrix integral technology \cite{Silvestrov:1990ha,  David:1992za,Hanada:2004im, Ishibashi:2005dh, Sato:2004tz, Ishibashi:2005zf, sss}, with ref. \cite{Ishibashi:2005zf} containing the result for general $(p',p)$.

This state of affairs is very reminiscent of the recent computations in the $c=1$ system, where the annulus amplitude between ZZ branes is also divergent, while the matrix side of the duality provides a finite unambiguous answer \cite{bryzz}.
It has been shown by one of us \cite{SenNormalization} that string field theory techniques allow us to compute $\mathcal{N}$ in this case and the result matches with the matrix computation.

The purpose of this note is to apply these string field theory tools to the $(2,p)$ minimal string theories and compute the value of $\mathcal{N}$ in these theories.
We find perfect agreement with the matrix integral computations \cite{Silvestrov:1990ha,  David:1992za,Hanada:2004im, Ishibashi:2005dh, Sato:2004tz, sss}.
We record the final result
\begin{align}
\mathcal{N} = 
T^{-\frac{1}{2}} \, \frac{i}{\sqrt{32\pi}}  \, \frac{\cot (\pi/p)}{\sqrt{p^2 - 4}}\, .
\label{Nanswer}
\end{align}
Let us make a few comments about the form of this answer.
First,  the combination $\mathcal{N} \, T^\frac{1}{2}$ is natural to consider since the dependence on $g_s$ cancels out in this combination.
This is important since it is impossible to fix the multiplicative constant between the genus counting parameters on the two sides of the duality, since we can always add the Euler characteristic term to the worldsheet action with an arbitrary coefficient.
So,  when trying to match precise numerical constants,  one should compute quantities that are independent of $g_s$,  like $\mathcal{N} \, T^\frac{1}{2}$ rather than $\mathcal{N}$ or $T$ separately.\footnote{
Another quantity like this would be the ratio of the disk amplitude to the square-root of the sphere amplitude \cite{msy}.  See also \cite{Alexandrov, Kazakov:2004du}.
}
On the matrix integral side,  the gaussian integral around the one-eigenvalue instanton gives a multiplicative factor in $\mathcal{N}$ that is proportional to $T^{-\frac{1}{2}}$.
On the string theory side,  this factor arises because the proper volume of the rigid $U(1)$ gauge group on the instanton is proportional to $T^{\frac{1}{2}}$ \cite{SenNormalization}.  
Division by this gauge group volume in the path integral produces the factor of
$T^{-\frac{1}{2}}$.
Second,  the overall sign of the right hand side of (\ref{Nanswer}) is ambiguous on both sides of the duality,  as it depends on a two-fold choice of the contour of integration over one unstable mode. 
One should make this choice so that the result is the same for the matrix integral and the string theory.
Third,  the normalization constant $\mathcal{N}$ is purely imaginary and the instanton correction we are studying computes the leading imaginary part of the free energy.
In this sense,  this correction is similar to the case of ``bounce" solutions in instanton physics \cite{Coleman} and the instanton correction is meaningful.
Finally, note that the coefficient on the right hand side of (\ref{Nanswer}) is finite in the JT gravity limit $p \to\infty$.

The organization of this paper is as follows.
In section \ref{sec:matrix},  we present the computation of $\mathcal{N}$ in the double-scaled one-matrix integral, which is dual to the $(2,p)$ minimal string.
The results of this section are not new, and we are including them to illustrate the relevant tools in the simpler setting of the one-matrix integral.
In section \ref{sec:strings},  we first present a general string field theory analysis of the divergences in the cylinder diagram with both boundaries lying on a D-instanton. 
We then apply these tools to the $(2,p)$ minimal string and obtain a finite answer that agrees with the matrix integral result.
In section \ref{sec:pprime},  we make a few remarks about the extension of these results to the more general $(p',p)$ minimal string.

\section{The matrix computation}
\label{sec:matrix}
In this section we will compute the normalization constant $\mathcal{N}$ for the one-matrix integrals that are dual to the $(2,p)$ minimal string.
The results in this section are not new and can be found in many papers,  including \cite{sss, Silvestrov:1990ha,  David:1992za,Hanada:2004im, Ishibashi:2005dh, Sato:2004tz,
Ishibashi:2005zf,Gregori:2021tvs}. We choose to follow the streamlined presentation given in the recent work \cite{sss}.

We start by explaining the setup.
The starting point is an integral over all $L \times L$ hermitian matrices
\begin{align}
\mathfrak{Z} = \int d H \, e^{- L\Tr V(H)}\, .
\label{defz}
\end{align}
Here $V$ is a potential which can be taken to be an even polynomial of degree $p+1$.
The matrix integral $\mathfrak{Z}$ is a function of the coefficients in this polynomial.
In the large $L$ limit,  we can talk about a smooth density of eigenvalues and it is supported on a finite interval on the real axis.  
The double-scaling limit refers to a procedure where,  in addition to taking $L \to \infty$,  we zoom in near the left edge of the spectrum and tune the coefficients of the potential such that the dominant double-line Feynman diagrams in the perturbation expansion of (\ref{defz}) resemble continuum surfaces \cite{DiFrancesco:1993cyw} .
In this limit, the density of states is non-normalizable and is supported on the entire positive real axis.

We focus on the so-called ``conformal background" \cite{Moore:1991ir},  where the leading density of states in the double-scaling limit reads\footnote{
To get to the density of states in JT gravity,  we need to take $\kappa \sim p^2$ as $p \to\infty$ \cite{sss}.
}
\begin{align}
\langle \rho (E) \rangle^{(0)} &= 
\frac{e^{S_0}}{\pi} \sinh \left( p \arcsinh  \sqrt{ \frac{E}{2\kappa}} \right)\, \Theta(E) \, .
\label{rhoconformal}
\end{align}
Here $\Theta(E)$ denotes the Heaviside theta function.
This is the density of states that is dual to standard Liouville theory with only the cosmological constant term in the action turned on.
See, for example, \cite{msy} for an explicit family of potentials that lead to the density of states (\ref{rhoconformal}) in the double scaling limit.
Here,  $e^{S_0}$ is the genus counting parameter after taking the double-scaling limit and $\kappa$ is an arbitrary energy scale.

Using the relationship between the form of the density of states and the spectral curve,  we conclude that the spectral curve is given by \cite{Seiberg:2003nm, seibergannulus}
\begin{align}
y(z)
= \sin \left( p\arcsin \frac{z}{\sqrt{2\kappa}} \right)
= (-1)^\frac{p-1}{2} \, T_p \left( \frac{z}{\sqrt{2\kappa}} \right)
\,  ,
\label{yconf}
\end{align}
where $T_p$ denotes the $p$-th Chebyshev-T polynomial.
One way to see this is to note that the leading density of states $\langle \rho (E) \rangle^{(0)} $ is determined from the spectral curve as $\langle \rho (E) \rangle^{(0)}  = -i \pi^{-1} e^{S_0} y(i \sqrt{E})$ for $E>0$.
It is also a standard result in one-matrix integrals that the derivative 
$V_\text{eff}'(E)$ of the one-eigenvalue effective potential $V_\text{eff}(E)$, that
includes contributions from both the potential $V$ that appears in \eqref{defz} and
the Vandermonde determinant, is proportional to $y(\sqrt{-E})$ in the forbidden region $E<0$ (see, for example, \cite{sss} for a recent exposition).
The precise relationship is
\begin{align}
V_\text{eff}'(E) = e^{S_0} \left(-2y \left(\sqrt{-E} \right) \right) \quad\quad (\text{for } E < 0)  
\, .
\label{veffy}
\end{align}
Integrating this using \eqref{yconf} and taking $V_\text{eff}(E=0) = 0$ we get,  for $E<0$ that
\begin{align}
V_\text{eff}(E) &= -2\, e^{S_0} \, \kappa
\left[ {1\over p+2} \sin\left( (p+2) \arcsin\sqrt{{-E\over 2\kappa}}\right) - {1\over p-2} \sin\left( (p-2) \arcsin\sqrt{{-E\over 2\kappa}}\right)\right] \label{veffexplicit}\\
&= 2 \, e^{S_0} \,  \kappa \, (-1)^\frac{p-1}{2} 
\left[
\frac{1}{p+2} \, T_{p+2}\left(\sqrt{\frac{-E}{2\kappa}}\right)
- \frac{1}{p-2} \, T_{p-2}\left(\sqrt{\frac{-E}{2\kappa}}\right)
\right]\, . \label{veffChebyshev}
\end{align}

Let us now look at the extrema of the one-eigenvalue effective action.
From (\ref{veffy}) and (\ref{yconf}),  we see that as we move towards negative energies starting at $E=0$, the first zero of $V_\text{eff}'(E)$ occurs at
\begin{align}
E^\star = - 2 \kappa \,   \sin^2 \frac{\pi}{p} \, .
\label{estar}
\end{align}
We record the values of $V_\text{eff}(E^\star)$ and $V_\text{eff}''(E^\star)$,  which are obtained from (\ref{veffexplicit}) and (\ref{veffy}) using (\ref{yconf}):
\begin{align}
V_\text{eff}(E^\star) &= e^{S_0} \kappa \,  \frac{4p\sin(2\pi/p)}{p^2 -4}  \, ,\label{veffestar}\\
V_\text{eff}''(E^\star) &= - e^{S_0} \kappa^{-1} \frac{p}{\sin (2\pi/p)}\, .\label{veffpp}
\end{align}

Now we organize various contributions to the integral (\ref{defz}) depending on how many eigenvalues are in the classically allowed region $E>0$ and how many are in the classically forbidden region $E < 0$.
The leading contribution $\mathfrak{Z}^{(0)}$ comes from the integration region where all eigenvalues are in the classically allowed region. 
The next important contribution $\mathfrak{Z}^{(1)}$ comes from the integration region when only one eigenvalue is in the forbidden region.
Next,  we borrow a couple of results from \cite{sss, David:1992za, Hanada:2004im, Sato:2004tz},  which in the notations of \cite{sss} are as follows:
\begin{align}
\frac{\mathfrak{Z}^{(1)}}{\mathfrak{Z}^{(0)}} &= \int_F d E\, \langle \rho(E) \rangle \, , 
\label{z1byz0}\\
\langle \rho(E) \rangle &= \frac{1}{-8\pi E} \exp (- V_\text{eff}(E))  \quad \text{for } E < 0\, .
\label{rhoeveff}
\end{align}
Here the subscript $F$ on the integral denotes integration over the classically forbidden region $E < 0$.
The formula (\ref{rhoeveff}) captures the small amount of quantum mechanical leakage of eigenvalues into the classically forbidden region.\footnote{
As commented upon in \cite{sss}, the expression for $V_\text{eff}$ in (\ref{veffChebyshev}) is negative for certain intervals on the negative real axis.  However, in the regime $E \in [E^\star,0]$,  with $E^\star$ as in (\ref{estar}),  this issue does not arise,  and this interval is all that we will need.  See the discussion of the integration contour below.
}

\begin{figure}
    \centering
    \includegraphics[width = 0.4\textwidth]{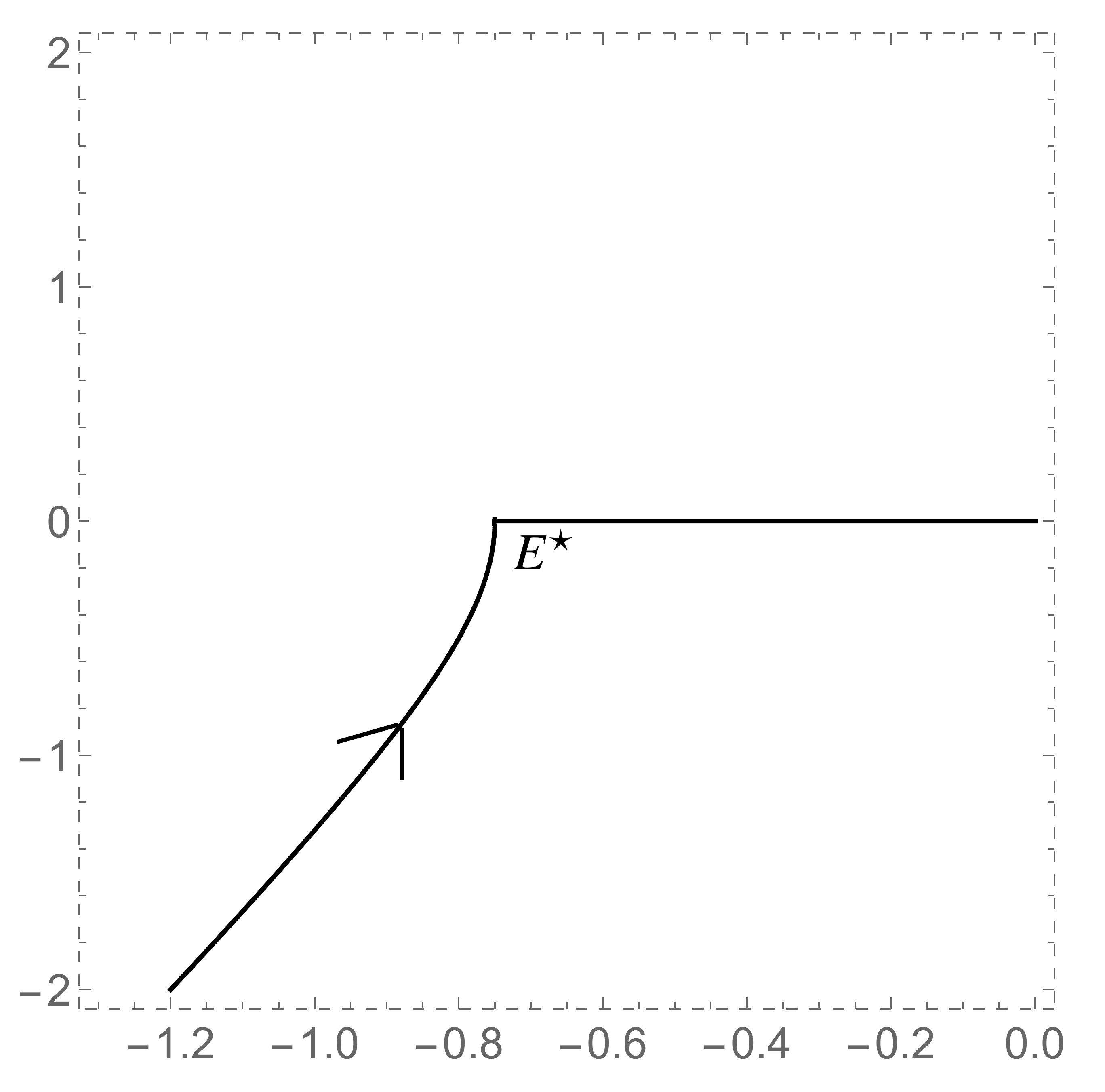}
    \caption{ The contour of integration for the eigenvalues showing that we only need to include ``half" of the steepest descent contour for the instanton saddle point.  We have shown the numbers for the case $p=3$ with $\kappa = \frac{1}{2}$ but it is qualitatively similar for all $p$.  In string theory,  the integration contour for the open string tachyon also looks like this.}
    \label{figcontour}
\end{figure}

We now plug in (\ref{rhoeveff}) into (\ref{z1byz0}) and use the saddle point approximation about $E^\star$ to compute the integral (along a contour to be specified momentarily):
\begin{align}
\frac{\mathfrak{Z}^{(1)}}{\mathfrak{Z}^{(0)}}
 &= 
\frac{1}{-8\pi E^\star} 
 \exp \left( - V_{\text{eff}}(E^\star)  \right)
\int d E \,  \exp \left[ 
		\frac{1}{2} \left \vert V_\text{eff}''(E^\star) \right\vert 
			(E-E^\star)^2  \right]\\
	&= \frac{1}{-8\pi E^\star} \exp \left( - V_{\text{eff}}(E^\star)  \right) \times \frac{i}{2} \sqrt{\frac{2\pi}{\vert V_{\text{eff}}''(E^\star) \vert}} \,  .
	\label{z1byz0int}
\end{align}
It is important to note from (\ref{veffpp}) that $V_\text{eff}''(E^\star) < 0$ and thus the steepest descent contour is parallel to the imaginary-$E$ axis.
Furthermore,  we only integrate over half of the steepest descent contour, since,  in the perturbative region $E \gg \kappa$,  the defining contour must lie along the real axis \cite{sss}.
Figure \ref{figcontour} shows this contour.
On the string theory side,  this ``unstable mode" is the open string tachyon and one has a similar contour of integration over the tachyon mode \cite{SenNormalization}.\footnote{
In fact, such a contour is common in decay rate computations using bounce solutions. 
See, for example,  \cite{Coleman}.
}
These facts give us the factor of $i/2$ in the gaussian integral.\footnote{
Since we are only interested in computing the imaginary part,  we don't need to worry about the part of the contour along the real axis,  which contributes something real.}

Comparing (\ref{z1byz0int}) to (\ref{zexpansion}) and using equations (\ref{estar}), (\ref{veffestar}) and (\ref{veffpp}),  we get
\begin{align}
T &= V_{\text{eff}}(E^\star) =  e^{S_0} \kappa \,  \frac{4p\sin(2\pi/p)}{p^2 -4}  \,  ,  \label{veffmatrixp} \\
\mathcal{N} &= 
e^{-\frac{S_0}{2}} \kappa^{-\frac{1}{2}} \,
\frac{i}{16\sqrt{\pi}} \, 
\sqrt{
\frac{\cos (\pi/p)}{p \sin^3 (\pi/p)}
}\, .
\end{align}
As explained in the introduction,  it is natural to factor out $T^{-\frac{1}{2}}$ from the expression for $\mathcal{N}$, and so we write the above result as
\begin{align} \label{ematrixfinal}
\mathcal{N} = T^{-\frac{1}{2}} \, \frac{i}{\sqrt{32\pi}}  \, \frac{\cot (\pi/p)}{\sqrt{p^2 - 4}}\, .
\end{align}
Refs. \cite{Silvestrov:1990ha, David:1992za, Hanada:2004im, Ishibashi:2005dh} contain this result for $p=3$, while 
the result for general $p$ can be found in \cite{Sato:2004tz}.\footnote{Note that some of these references are computing an integral over the full steepest contour through the saddle point,  and others are including contributions from both ends of the eigenvalue cut, and thus the pre-factors quoted there are a multiple of the value in (\ref{ematrixfinal}\label{foot:fact}). 
}
Ref. \cite{sss} was interested in the limit $p \to \infty$.

We would like to explain one subtlety in the above analysis.
One can explicitly check that the effective potential given in equation (\ref{veffChebyshev})
has $(p-1)/2$ extrema on the negative-$E$ axis.
Roughly half of them are maxima and half are minima.
The extremum at $E^\star$ in (\ref{estar}) is the one closest to the origin and is a local maximum.
However,  even among the local maxima,  this is not the one with the smallest value of the effective potential, in general.
This raises the question of why we have chosen the saddle point $E^\star$ in (\ref{estar}) as the relevant saddle.
The point is that we want the perturbation series of the matrix integral to match with the vacuum string perturbation theory,  and so we should not allow the integration contour for the matrix eigenvalues to pass through regions on the real axis with $V_\text{eff} < 0$,  since these regions will give real contributions to the matrix integral that are much larger than the terms in perturbation theory around the saddle point (\ref{rhoconformal}). 
This can be avoided by turning the integration contour
along the steepest descent contour once it reaches $E^*$.

\section{The string theory computation}
\label{sec:strings}

In this section we shall describe the string theory computation of the leading imaginary
part of the partition function, arising from a single ZZ-instanton contribution.

The string theory that is dual to the double-scaled one-matrix integral described in section \ref{sec:matrix} is Liouville theory coupled to the $(2,p)$ minimal model and the $bc$-ghost system.
The $b$ parameter that appears in the Liouville lagrangian is determined by $p$ and is such that the total central charge of Liouville,  the matter CFT and ghosts adds up to zero.
One finds $b = \sqrt{2/p}$.

\subsection{The cylinder diagram and its divergences} \label{sintrocyl}

We shall begin by describing some general issues that arise in the analysis
of the cylinder diagram with boundaries lying on a D-instanton (whose analog in non-critical string theory is the ZZ instanton).
We can express the cylinder partition function in the open string channel as:
\be\label{ecylinder}
A = \int_0^\infty{dt\over 2t} F(t)\, ,
\ee
where $F(t)$ has the structure
\be
F(t) =\sum_b e^{-2\pi h_b t} - \sum_f e^{-2\pi \hat h_ f t}\, ,
\ee
$h_b$ and $\hat h_f$ being the $L_0$ eigenvalues of the bosonic and fermionic states of the open string with any ghost number and subject to the Siegel gauge condition.
A state $\ket{\chi}$ is said to satisfy the Siegel gauge condition if
\be\label{esiegel}
b_0|\chi\rangle=0\, \quad (\text{Siegel gauge condition}).
\ee
The states are taken to be fermionic if they carry even ghost number and bosonic if they carry odd ghost number -- this is the correct assignment of statistics when we regard the coefficients of these states as modes of the open string field on the D-instanton.
The Siegel gauge condition \refb{esiegel} is needed,  since without this condition there will be an equal number of bosonic and fermionic states
related by the action of the ghost zero modes $b_0$ or $c_0$,  and the partition function
will vanish. 
The way this gets implemented in the worldsheet computation is via the insertion of $b_0 c_0$ to soak up the ghost zero modes on the cylinder \cite{Polchinski:1998rq}.

In theories of interest to us in this paper,  the integral \refb{ecylinder} has no divergence in the $t\to 0$ limit, indicating that the (regulated) number of fermionic and bosonic states are equal. 
In the hypothetical situation where $h_b$ and $\hat h_f$ are all positive,  there are no divergences in the $t\to \infty$ limit either, and $A$ is given by
\be \label{eproduct}
A = {1\over 2}\ln {\prod_f \hat h_f\over \prod_b h_b}\, .
\ee
For positive $h_b,\hat h_f$
this can be used to express the normalization factor $\mathcal{N}$ accompanying the instanton amplitude as an integral,
\begin{align}
\mathcal{N} = e^A =
\left({\prod_f \hat h_f\over \prod_b h_b}\right)^{\frac{1}{2}} &=
\frac{{\prod'_f} \hat h_f}{\prod_b h_b^{1/2}} 
= \int \prod_b {d \phi_b\over \sqrt{2\pi}}\, {\prod_f}'dp_f dq_f \, \exp\left[
-{1\over 2} \sum_b h_b \phi_b^2 - {\sum_f}' \hat h_f p_f q_f\right]\, ,
\label{epathintegral}
\end{align}
where $\phi_b$ are grassmann even variables and $p_f,q_f$ are grassmann odd variables.
The prime on the summation and the product symbols indicate that, since $\hat h_f$'s
occur in pairs,\footnote{This can be seen as follows. For any choice of basis
states $\{|a\rangle\}$ for Siegel gauge states with a fixed $L_0$ eigenvalue, 
$\langle a|c_0|b\rangle$ gives a non-degenerate inner product matrix. Since this inner
product pairs states of ghost number $n$ and $(2-n)$, we see that for every $n$ other
than $n=1$, the $L_0$ eigenvalues occur in pairs in sectors with ghost numbers $n$
and $(2-n)$. Since fermions arise from even
ghost number sector,  the $\hat h_f$'s always occur in pairs. 
}
we let the sum and product over $f$ run over half the number of original variables,  and for each $f$ introduce a pair of grassmann odd variables $p_f,q_f$. 
The final expression in \refb{epathintegral} may be regarded as the path integral over open string fields  in Siegel gauge,  with the understanding that open string fields live on the zero dimensional worldvolume of the D-instanton and therefore are just ordinary variables.
See appendix \ref{conventions} for our conventions for the open string field action.

As long as $h_b$ and $\hat h_f$ are positive,  \refb{ecylinder}, \refb{eproduct} and \refb{epathintegral} are all well defined and are identically equal. 
However in most situations,  some of the $h_b$'s are negative or zero,  and some of the $\hat h_f$'s may vanish. 
In that case \refb{ecylinder} and \refb{eproduct} are ill-defined. 
The final expression in \refb{epathintegral} is also ill-defined but we can try to make sense of this using insights from string field theory. 
We shall now describe this procedure. 

First we note that, for $h_b,  h_f>0$,  we can pick any non-negative integer $n$ and write hybrid expressions for $A$ and $\mathcal{N}$ as
\begin{align}
A &= \int_0^\infty{dt\over 2t} \left[ F(t) - \sum_{b=1}^{2n} e^{-2\pi t h_b} 
+ \sum_{f=1}^{2n} e^{-2\pi t \hat h_f}\right] + \int_0^\infty{dt\over 2t} \left[
\sum_{b=1}^{2n} e^{-2\pi t h_b} - \sum_{f=1}^{2n} e^{-2\pi t \hat h_f} \right] \non\\
&= \int_0^\infty{dt\over 2t} \left[ F(t) - \sum_{b=1}^{2n} e^{-2\pi t h_b} + \sum_{f=1}^{2n} e^{-2\pi t \hat h_f}\right] + {1\over 2} \log\left[
{\prod_{f=1}^{2n} \hat h_f\over \prod_{b=1}^{2n} h_b}\right], \\
\mathcal{N} &=  e^A = \exp\left[\int_0^\infty{dt\over 2t} \left[ F(t) - \sum_{b=1}^{2n} e^{-2\pi t h_b} + \sum_{f=1}^{2n} e^{-2\pi t \hat h_f}\right]\right] 
\non\\ &  \hspace{50pt} \times
\int \prod_{b=1}^{2n} {d\phi_b\over \sqrt{2\pi}} \prod_{f=1}^n dp_f dq_f \exp\left[-{1\over 2} \sum_{b=1}^{2n}
h_b \phi_b^2 - \sum_{f=1}^n \hat h_f p_f q_f\right].\label{e1}
\end{align}
Now,  when some of the $h_b$'s or $\hat h_f$'s are negative or zero,  we shall choose $n$ to be such that for $b,f>2n$ all the $h_b$'s and $\hat h_f$'s are positive. 
Then the term in the first line of (\ref{e1}) is finite since we have subtracted the `bad' contributions involving $h_b,\hat h_f\le 0$ terms from $F(t)$. 
Furthermore, since the subtraction term vanishes as $t\to 0$,  the integral is free of divergences from the $t\to 0$ end as well.
Thus,  we are left with the goal of making sense of the integral over the modes $\phi_b$ for $b\le 2n$ and $p_f,q_f$ for $f\le n$.

\begin{table}
\centering
   \begin{tabular}{|c|c|c|c|c|c|}
   \hline
   State & $L_0$ eigenvalue & Ghost number & In Siegel gauge? & Field name & Grassmann parity \\
   & & & & & of field\\
   \hline
   $c_1 \ket{0}$ & $-1$ & 1 & Yes & $\phi_1$ & even\\
   $c_0 c_1 \ket{0}$ & $-1$ & 2 & No & - & odd\\
   $\ket{0}$ & $0$ & 0 & Yes &  $p_1 $ & odd \\
   $c_0 \ket{0}$ & $0$ & 1 & No & $\psi$  & even \\
   $c_{1} c_{-1} \ket{0}$ & $0$ & 2 & Yes & $q_1$ & odd \\
   $c_0 c_{1} c_{-1} \ket{0}$ & $0$ & 3 & No & -  & even\\
   \hline
   \end{tabular}
    \caption{ A list of states that are relevant for the discussion of divergences in the cylinder diagram. We have ordered the states first by their $L_0$ eigenvalues and then by their ghost numbers.  A state and the corresponding field appear multiplied together in the expansion of the open string field as $\ket{\Psi} = \phi_1 c_1 \ket{0} + \ldots$.}
    \label{tablestates}
\end{table}

For the D-instantons that we shall discuss,  the bad modes consist of one bosonic mode -- the tachyon mode $\phi_1$ corresponding to the state $c_1|0\rangle$ with $h_b=-1$,  and a pair of fermionic modes $p_1,q_1$ corresponding to  the states $i|0\rangle$ and $i c_1 c_{-1}|0\rangle$ with $\hat h_f=0$. 
The coefficients $i$ in these states have been chosen to ensure that the modes multiplying these states are real.
Since there is only one bad bosonic mode and two bad fermionic modes,  we can choose $n=1$ in \refb{e1}.
See Table \ref{tablestates} for a list of the states that are relevant for the discussion and their basic properties.

First we shall discuss the integration over the bosonic modes $\phi_1$ and $\phi_2$.
Since $h_2>0$, the integration over $\phi_2$ gives a standard gaussian integral. 
The integration over $\phi_1$ is problematic since the exponent takes the form $\exp(\phi_1^2/2)$. 
We shall carry out this integral by regarding this as a contour integral in the complex $\phi_1$ plane as follows\cite{SenNormalization}.
Since the vacuum without any D-instanton is represented by a particular solution of the open string field theory corresponding to some positive value $\beta$ of $\phi_1$, the integration contour must pass through $\beta$.  
For this reason we take the integration contour to lie along the positive real axis for $\Re(\phi_1)>0$. However once we reach $\phi_1=0$, we take the contour to be along (half of) the steepest descent contour -- either from $-i\infty$ to 0,  or from $i\infty$ to 0. 
The integration along the real axis is real and can be regarded as the perturbative contribution since the contour passes through the perturbative vacuum.
The leading imaginary part comes from the part of the contour from $\pm i\infty$ to 0. 
These two choices differ by a sign -- an ambiguity that is also present in the matrix model. 
Choosing the contour to be from $-i\infty$ to 0 for definiteness,  we can write (the
leading imaginary part of) the bosonic part of the integral as:
\be\label{e3.9}
\int_{-i\infty}^0 {d\phi_1\over \sqrt {2\pi}} 
e^{\phi_1^2/2} \int_{-\infty}^\infty {d\phi_2\over \sqrt {2\pi}} e^{-h_2\phi_2^2/2} 
= {i\over 2} h_2^{-1/2}\, .
\ee

Next we turn to integration over the fermion zero modes $p_1,q_1$. 
We can get physical insight into the origin of these modes if, instead of a D-instanton, we consider a D$p$-brane extending along some directions in space-time in any bosonic string theory. 
In that case the gauge field $a_\mu(k)$ living on the brane appears in the expansion of the open string field as a term proportional to $\int d^{p+1}k \, a_\mu(k) \alpha^{\mu}_{-1}c_1|k\rangle$ where $\alpha^\mu_n$ are the oscillators associated with the scalars $X^\mu$ describing coordinates tangential to the brane and $|k\rangle=e^{ik\cdot X(0)}|0\rangle$ are momentum carrying states. 
In string field theory,  gauge transformations appear as BRST exact states $Q_B \ket{\Lambda}$ (plus higher order terms),  and $\ket{\Lambda}$ is referred to as the ``gauge transformation parameter".
For instance,  usual spacetime gauge transformations of the gauge field 
$\delta a_\mu(k) \propto i k_\mu \theta(k)$ appear via the term $i\int d^{p+1}k \, \theta(k) |k\rangle$ in $\ket{\Lambda}$.      
Note that this term in $\ket{\Lambda}$ has ghost number zero.
Then the linearized gauge transformation $Q_B \ket{\Lambda}$ produces a term proportional to $i\int d^{p+1}k \,\theta(k) k_\mu  \alpha^{\mu}_{-1}c_1|k\rangle$. 
Comparing to the state representing the gauge field, we see that this generates the usual gauge transformation law $\delta a_\mu(k) \propto i\, k_\mu \theta(k)$.

The gauge transformation $Q_B \ket{\Lambda}$ also produces a state proportional to
$i\int d^{p+1}k \,\theta(k) k^2 c_0|k\rangle$. 
This translates to a transformation $\delta\psi(k)\propto k^2\theta(k)$ where $\psi(k)$ is the field multiplying the state $ic_0|k\rangle$.
The Siegel gauge choice corresponds to setting $\psi(k) = 0$.
This produces a Jacobian proportional to $k^2$,  which is represented by a pair of Fadeev-Popov ghosts $p_1(k)$, $q_1(k)$ multiplying the states $ic_1 c_{-1}|k\rangle$ and $i |k\rangle$.
Since these states have conformal weight $\hat h_1=k^2$,  integration over
$p_1$ and $q_1$ will precisely produce the required Fadeev-Popov determinant $k^2$, for $k\ne 0$.

Now,  the issue is that, on a D-instanton we have $k=0$. 
Thus,  neither the ``gauge field" nor the field $\psi$ multiplying $ic_0|0\rangle$ transforms, showing that the Siegel gauge choice breaks down. 
This is a reflection of the fact that the usual local U(1) symmetry on the D$p$-brane becomes a rigid symmetry on the D-instanton. 
The remedy is to go back to the ``original" form of the
path integral where we carry out integration over all the ``classical" modes of the theory and explicitly divide by the volume of the gauge group.
In string field theory language,  fields that multiply states of ghost number one are referred to as classical since the physical open string states belong to this sector.
Concretely,  among the states in table \ref{tablestates},  this means that instead of integrating over $\{\phi_1,  p_1,  q_1\}$,  we integrate over $\{\phi_1, \psi\}$ and divide by the volume of gauge group.
The precise normalization of the integration measure can be fixed by carefully following the line of argument described above and gives the replacement rule\cite{SenNormalization}:
\be\label{e3.10}
\int dp_1 dq_1\ \quad \longrightarrow \quad
\frac{\int d\psi \, e^{-\psi^2}}{ \int d\theta} = \frac{\sqrt \pi}{\int d\theta} \, .
\ee
The $-\psi^2$ in the exponent is the result of evaluating the open string field theory action for the out-of-Siegel-gauge grassmann-even mode $\psi$,\footnote{
The generalization of this term to a D$p$-brane would be $-\int d^{p+1}x \, (\psi + \gamma \, \partial_\mu a^\mu)^2$,  where the constant $\gamma$ is chosen such that the combination $\psi + \gamma \, \partial_\mu a^\mu$ is gauge invariant. }  
see appendix \ref{conventions}.
The quantity $\theta$ can be related to the rigid U(1) symmetry paramater $\tilde\theta$, under which an open string with one end on the instanton picks up a phase $e^{i\tilde\theta}$,  by comparing the string field theory gauge transformation to the rigid U(1) transformation. 
For canonically normalized fields and gauge transformation parameters, the transformation law of a charged field $\Phi$ is proportional to $i g_o \, \theta \, \Phi$,  as in conventional quantum field theories.
Here $g_o$ is defined to be the coefficient of the cubic term in the open string field theory action,  with conventions as described in appendix \ref{conventions}.
This should be equated to the transformation law $\delta\Phi=i\, \tilde\theta\, \Phi$ under infinitesimal rigid U(1) transformation. 
A detailed calculation of the constant appearing in the string field theory gauge transformation law leads to $\theta = \tilde\theta/g_o$\cite{SenNormalization}. 
On the other hand, the open string coupling $g_o$ is related to the instanton action $T$ via 
$g_o = (2\pi^2 T)^{-\frac{1}{2}}$.\footnote{
There are many ways to derive this result,  but the one that holds universally is from the observation that the tachyon vacuum solution in open string field theory has
action $T-(2\pi^2 g_o^2)^{-1}$ \cite{Sen:1999xm,Schnabl:2005gv}.    
Since this describes the vacuum,  we must equate this to zero,  leading to $T=(2\pi^2g_o^2)^{-1}$.
}
Therefore we
have
\be\label{e3.11}
\int d\theta = g_o^{-1} \int d\tilde\theta = 2\pi g_o^{-1}=2^\frac{3}{2} \pi^2 T^{\frac{1}{2}}\, ,
\ee
since $\tilde\theta$ has period $2\pi$. 

Substituting \refb{e3.9}, \refb{e3.10} and \refb{e3.11} into \refb{e1} we get:
\be\label{e.7}
\mathcal{N} = \exp\left[\int_0^\infty{dt\over 2t} \left[ F(t) - 
\left(e^{2\pi t} + e^{-2\pi h t} -2\right) \right]\right] \times i\, 
2^{-\frac{5}{2}} \pi^{-\frac{3}{2}} \, h^{-\frac{1}{2}} \, T^{-\frac{1}{2}},
\ee
where $h = h_2$.
One can easily check that the expression is independent of $h$ by taking derivative
with respect to $h$. 
Therefore we do not need to choose $h=h_2$, any choice
of $h>0$ will give the same result.

\subsection{Specialization to minimal string theory}
\label{sec:minimalstring}
We shall now use \refb{e.7} to compute the normalization of the instanton amplitude in
the $(2,p)$ minimal string theory.
The form of the integrand $F(t)$ for the cylinder diagram in minimal string theory is well-known \cite{Martinec:2003ka}.
Since we are studying the cylinder diagram,  we need to specify  boundary conditions for the worldsheet fields.
For the Liouville CFT,  we pick the ``$(m,n) = (1,1)$" ZZ boundary condition \cite{zz},  as this is the one that corresponds to the saddle point $E^\star$ in equation (\ref{estar}) in the matrix integral \cite{seibergannulus, Alexandrov}.
For the matter CFT,  we pick the Cardy state on both ends so that the open string channel only contains the identity character \cite{Cardy89}.\footnote{For the $(2,p)$ minimal string,  there are $(p-1)/2$ possible ZZ brane boundary conditions \cite{Seiberg:2003nm}.  
By comparing the relative tensions of these branes (given in, for example, \cite{seibergannulus}), to the relative heights of the extrema of the matrix effective potential (\ref{veffChebyshev}),  one can establish that it is the $(m,n) = (1,1)$ ZZ brane,  with identity character from the matter CFT,  that corresponds to  the matrix saddle point at $E^\star$ in (\ref{estar}) with $V_\text{eff}(E^\star)$ as in (\ref{veffmatrixp}).
}

Let $t$ be the modulus of the cylinder that corresponds to time in the open string channel and let $q = e^{-2\pi t}$.
The partition function of Liouville theory on the cylinder with $(m,n) = (1,1)$ ZZ boundary conditions on both ends is \cite{zz}
\begin{align}
Z_\text{Liouville} (t)
&= \left( q^{-1} -1 \right) q^{-\frac{1}{4} (b^{-1}-b)^2 }  \, \eta(i t)^{-1}
= \left( q^{-1} -1 \right) q^{-\frac{(p-2)^2}{8p} }  \, \eta(i t)^{-1} \, .
\end{align}
The partition function of the matter CFT with the given boundary conditions equals the identity character in the minimal models,  which is given by \cite{Rocha-Caridi,  DiFrancesco:1997nk}
\begin{align}
Z_\text{matter}(t)  &= \eta(i t)^{-1} \sum_{k = - \infty}^\infty \left( 
q^{\frac{(4pk + p  - 2 )^2}{8p}}
-
q^{\frac{(4pk + p +2 )^2}{8p}} 
\right)\, .
\end{align}
Multiplying the contribution $\eta(i t)^2$ from the ghosts (see, for example, \cite{Polchinski:1998rq}),  we find
\begin{align} \label{e.14}
F(t) = \left( e^{2\pi t} - 1\right) \sum_{k = -\infty}^{\infty} \left(
e^{-2\pi t k (2p k + p - 2)}
- 
e^{-2\pi t (pk+1)(2k + 1)}
\right).
\end{align}
It is important to note that the leading terms in $F(t)$ as $t \to \infty$ are the ones with $k=0$:
\begin{align}
F(t) = \left(e^{2\pi t}-1\right) \left( 1 - e^{-2\pi t} + O(e^{-4\pi t}) \right)
= e^{2\pi t} - 2 + O(e^{-2\pi t})\, .
\end{align}
As already discussed in section \ref{sintrocyl},  the $e^{2\pi t}$ term arises from the open string tachyon,  while the $-2$ arises from the two ghost zero modes.\footnote{
In the $c=1$ case,  we have $F(t) = e^{2\pi t} - 1$ exactly.  The change of coefficient in the $L_0 = 0$ sector comes from an additional bosonic zero mode that corresponds to time translations of the D-instanton \cite{bryzz, SenNormalization}.
}

If we substitute \refb{e.14} into \refb{e.7} and choose $h=1$, we can see that the $k=0$ term
in the sum exactly cancels the subtraction term $\left(e^{2\pi t} + e^{-2\pi  t} -2\right)$.
The rest of the terms may be analyzed using the general result:
\be
\int_0^\infty {dt\over 2t} \left(e^{-2\pi h_1 t}- e^{-2\pi h_2 t}\right) = {1\over 2}\ln {h_2\over h_1}\,.
\ee
Using this we can rewrite \refb{e.7} as
\begin{align}
\mathcal{N} &= i\, 2^{-\frac{5}{2}} \pi^{-\frac{3}{2}} T^{-\frac{1}{2}} 
\prod_{k=-\infty\atop k\ne 0}^\infty\left[ {(pk+1)(2k+1)-1 \over
k(2pk+p-2)-1} \, {k(2pk+p-2)\over (pk+1)(2k+1)}
\right]^{\frac{1}{2}} \non\\
&= 
i\, 2^{-\frac{5}{2}} \pi^{-\frac{3}{2}} T^{-\frac{1}{2}} 
\left[\prod_{k=-\infty\atop k\ne 0}^\infty
\frac{1 - \frac{4}{p^2(2k+1)^2}}{1 - \frac{1}{p^2 k^2}}\right]^{\frac{1}{2}}\, .
\label{efinal}
\end{align}
We now use
\begin{align}
\sin \pi x &= \pi x \prod_{k=1}^\infty \left(1 - \frac{x^2}{k^2} \right) \, , \quad 
\cos \pi x = \frac{\sin 2\pi x}{2 \sin \pi x} = \prod_{k=1}^\infty \left(1 - \frac{4x^2}{(2k-1)^2} \right)
\label{ecosineprod}
\end{align}
to write
\be
\cot \frac{\pi}{p} = \frac{p}{\pi} \prod_{k=1}^\infty
\frac{1 - \frac{4}{p^2(2k-1)^2}}{1 - \frac{1}{p^2 k^2}}\, .
\ee
Using this it is easy to see that product over terms in \refb{efinal} for $k<0$ produces
$ \frac{\pi}{p}\cot \frac{\pi}{p}$.
For positive $k$,  (\ref{efinal}) is missing the $1-4/p^2$ term from the cosine infinite product in \refb{ecosineprod}. 
Thus the infinite product term in \refb{efinal} produces
\begin{align}
 \frac{\pi}{p}\cot \frac{\pi}{p} \times \frac{1}{1 - \frac{4}{p^2}}  \times  \frac{\pi}{p}\cot \frac{\pi}{p}
 = \frac{\pi^2}{p^2 - 4} \cot^2 \frac{\pi}{p}\, .
\end{align}
Using this result in (\ref{efinal}) yields
\be \label{efinala}
\mathcal{N} = 
T^{-\frac{1}{2}} \, \frac{i}{\sqrt{32\pi}}  \, \frac{\cot (\pi/p)}{\sqrt{p^2 - 4}}\, ,
\ee
in perfect agreement with the matrix integral result \refb{ematrixfinal}.

\section{Generalization to $(p',p)$ models}
\label{sec:pprime}
A more general class of minimal string models is Liouville theory coupled to the $(p',p)$ minimal model and the $bc$-ghost system.
Here $p$ and $p'$ are relatively-prime integers with $p > p' \geq 2$.
The $b$ parameter of Liouville theory is determined by the requirement that the total central charge vanishes; the result is $b = \sqrt{p'/p}$.

The Liouville sector admits ZZ boundary conditions labeled by two integers $(m,n)$ \cite{zz}.
We leave a general analysis to future work, but for illustration purposes, we note here that the computation in section \ref{sec:minimalstring} can be extended to the $(p',p)$ minimal string with the same boundary conditions. 
That is,  we pick the $(m,n) = (1,1)$ ZZ state for Liouville,  and for the matter CFT,  we pick the Cardy state on both ends so that the open string channel only contains the identity character \cite{Cardy89}. 
This gives the partition functions \cite{zz, Rocha-Caridi,  DiFrancesco:1997nk}
\begin{align}
Z_\text{Liouville}(t)
&= \left( q^{-1} -1 \right) q^{-\frac{(p-p')^2}{4pp'} }  \, \eta(i t)^{-1} \,  , \\
Z_\text{matter}(t)  &= \eta(i t)^{-1} \sum_{k = - \infty}^\infty \left( 
q^{\frac{(2pp'k + p - p' )^2}{4pp'}}
-
q^{\frac{(2pp'k + p + p' )^2}{4pp'}} 
\right)\, .
\end{align}
Combining the Liouville, matter and ghost contributions to $F(t)$,  using (\ref{e.7}),  and following the steps in section \ref{sec:minimalstring}, we get
\begin{align}
\mathcal{N} &=  i\, 2^{-\frac{5}{2}} \pi^{-\frac{3}{2}} T^{-\frac{1}{2}} 
\prod_{k=-\infty\atop k\ne 0}^\infty\left[ {(pk+1)(p'k+1)-1 \over
k(pp'k+p-p')-1} \, {k(pp'k+p-p')\over (pk+1)(p'k+1)}
\right]^{1/2} \non\\
&= i\, 2^{-\frac{5}{2}} \pi^{-\frac{3}{2}} T^{-\frac{1}{2}} \prod_{k=1}^\infty
\left[ 
\left(1 - \frac{1}{k^2}\left( \frac{1}{p} - \frac{1}{p'}\right)^2 \right) 
\left(1 - \frac{1}{k^2}\left( \frac{1}{p} + \frac{1}{p'}\right)^2 \right) 
\left(1 - \frac{1}{k^2 p^2} \right)^{-2}
\left(1 - \frac{1}{k^2 p'^2} \right)^{-2}
\right]^{\frac{1}{2}} \non\\
&= T^{-\frac{1}{2}}\,\frac{i}{\sqrt{32\pi}} \,  \left[
\frac{\sin\left(\frac{\pi}{p} + \frac{\pi}{p'} \right)
\sin\left(\frac{\pi}{p'} - \frac{\pi}{p} \right)}
{\sin^2(\pi/p) \sin^2(\pi/p') \, (p^2 - p'^2)}\right]^\frac{1}{2}
= T^{-\frac{1}{2}} \,  \frac{i}{\sqrt{32\pi}} \sqrt{\frac{\cot^2(\pi/p) - \cot^2(\pi/p')}{p^2 - p'^2}}\, .
\label{ppprime}
\end{align}
This agrees with (\ref{efinala}) when $p'=2$.

For $p>p'\geq 3$,  these string theories are dual to the double-scaled limit of a two-matrix integral \cite{Douglas:1990pt, Daul:1993bg}.
The two-matrix integral is more complicated, so we won't go into the full analysis of the eigenvalue instanton in this case \cite{Kazakov:2004du, Ishibashi:2005zf},  and just note that the result (\ref{ppprime}) agrees with the $m=n=1$ expression given in \cite{Ishibashi:2005zf}.\footnote{In carrying out this comparison, following the comment in footnote \ref{foot:fact}, we have divided the result of \cite{Ishibashi:2005zf} by two. 
However, since the saddle point corresponding to $m=n=1$ is not the dominant saddle point in general, we need to carefully analyze the full integration contour to figure out how the steepest descent contour fits in.  
Since this issue exists both in the matrix model and in string theory,  we expect any additional factor to affect both sides in the same way.  Hence it should not affect the comparison.}
We leave a fuller investigation of the two-matrix case to future work.

\paragraph{Acknowledgments.} 
We would like to thank D. Stanford for suggesting this problem and for a careful reading of the manuscript.
D.S.E. would like to acknowledge the Shoucheng Zhang Graduate Fellowship for support.
R.M.  is supported in part by Simons Investigator Award \#620869 and by AFOSR grant FA9550-16-0092. 
C.M. is supported in part by the U.S. Department of Energy, Office of Science, Office of High Energy Physics under QuantISED Award DE-SC0019380 and contract DE-AC02-05CH11231.
A.S. is supported by ICTS-Infosys Madhava Chair Professorship and the J. C. Bose fellowship of the Department of Science and Technology, India.

\appendix
\section{Conventions for the open string field theory action}
\label{conventions}
Let us denote by $\ket{\Psi}$ the open string field which takes the form
\begin{align}
\ket{\Psi} = \phi_1 c_{1} \ket{0} + i\psi c_0 \ket{0} + \ldots
\end{align}
The vacuum state is normalized so that
\begin{align}
\bra{0} c_{-1} c_0 c_{1} \ket{0}  = 1\, .
\end{align}
Our starting point is the path integral over fields with ghost number one,  divided by the volume of the gauge group.
In the string field theory literature,  the fields with ghost number one are known as ``classical" fields since the physical open string states belong to this sector.
We take the weight in the path integral to be $\exp(-S)$ with the quadratic part of the action being
\begin{align} 
S = \frac{1}{2} \bra{\Psi} Q_B \ket{\Psi} \, .
\label{sftaction}
\end{align}
The BRST charge $Q_B$ is given by
\begin{align}
Q_B &= \oint  \frac{d z}{2\pi i} \, (c \, T_m +  \colon \hspace{-3pt}b \, c \, \partial c \colon\hspace{-3pt} ) \, ,
\end{align}
where $T_m$ is the matter stress tensor. There is also a cubic term in the action\cite{Witten:1985cc}. 
See  for example, \cite{Sen:1999xm} for a detailed form of this coupling.
If we normalize the string field so that the kinetic term is independent of the coupling
as in \eqref{sftaction},
then the cubic term has an explicit factor of the open string coupling $g_o$.

From the above equations,  one can see, for example, that the contribution of the tachyon field $\phi_1$ to the quadratic action is
\begin{align}
S \supset
\frac{1}{2} \bra{0} \phi_{1} c_{-1} \, Q_B \, \phi_1 c_1 \ket{0} = - \frac{1}{2} \phi_1^2\, ,
\end{align}
and thus the weight in the path integral is $\exp(\phi_1^2/2)$. 
The action for $\psi$ is similarly seen to be $\psi^2$.

\bibliographystyle{apsrev4-1long}
\bibliography{zz_annulus}
\end{document}